\documentclass[10pt]{article}

\usepackage{bm,graphicx}

\usepackage[english]{babel}

\usepackage{epsfig, amssymb, amsthm, amsmath}

\newcommand{\finpr}{\hfill $\square$ \vspace{2mm}}

\def\be{\begin{eqnarray}}
\def\ee{\end{eqnarray}}
\def\bee{\begin{eqnarray*}}
\def\eee{\end{eqnarray*}}
\newtheorem{thm}{Theorem}

\newtheorem{lem}{Lemma}

\begin{document}

\title{\bf Quantum matchgate computations \\ and linear threshold gates}

\author{Maarten Van den  Nest \\ \\ {\normalsize\it Max-Planck-Institut f\"ur Quantenoptik,} \\ {\normalsize\it  Hans-Kopfermann-Stra\ss e 1}, {\normalsize \it D-85748 Garching, Germany. }}


\maketitle

\begin{abstract}
The theory of matchgates is of interest in various areas in physics and computer science. Matchgates occur in e.g. the study of fermions and spin chains, in the theory of holographic algorithms and in several recent works in quantum computation.
In this paper we  completely characterize the class of boolean functions computable by unitary two-qubit matchgate circuits with some probability of success. We show that this class precisely coincides with that of the \emph{linear threshold gates}. The latter is a fundamental family which appears in several fields, such as the study of neural networks. Using the above characterization, we further show that the power of matchgate circuits is surprisingly trivial in those cases where the computation is to succeed with high probability. In particular,  the only functions that are matchgate-computable with success probability greater than $\frac{3}{4}$ are functions  depending on only a \emph{single} bit of the input.
\end{abstract}


\section{Introduction}

One of the great virtues of the field of quantum computation is that it interconnects fundamental questions in physics and computer science.  The concept of the quantum computer \cite{De89} precisely captures the intrinsic computational power locked within quantum mechanics \cite{Be97}, and makes it possible to address deep problems such as the relationship between quantum and classical computational capabilities \cite{Go98, Jo02, Va02}. At the same time, within the theory of quantum computation it is possible to characterize, in a precise sense, how ``hard'' is it to simulate physical systems of interest, such as certain ground state problems \cite{Ki99} and time evolutions \cite{Ll96}.

Of particular interest, also in recent work, is the class of quantum processes generated by \emph{matchgates} \cite{Va02,Di04,Ca06,Jo08,Br09,Jo09,Fi61}. The latter are a class of unitary two-qubit operations that are defined by certain algebraic constraints.
The theory of matchgates is an instance of a research area that displays strong connections to both physics and computer science \cite{Va02,Di04,Ca06,Jo08,Br09,Jo09,Fi61}. In the study of strongly correlated systems, for example, the dynamics of an important class of 1D quantum systems such as the XY model are modeled by matchgate circuits i.e. for such hamiltonians $H$  one can construct a poly-size matchgate circuit ${\cal C}_t$ such that ${\cal C}_t = e^{itH}$ for any time $t$ (see e.g. \cite{Jo08}). Employing mappings between spin-$\frac{1}{2}$ systems and fermions, matchgate circuits further describe the dynamics of all non-interacting fermionic systems \cite{Di04}. In the theory of  quantum computation, matchgates are of particular interest as they provide a key example of class of nontrivial quantum circuits that \emph{cannot} offer any speed-up over classical computers (in spite of e.g. the complex entangled states such circuits may generate) \cite{Va02, Jo08}. In addition, matchgate computations were recently found to be equivalent to space-bounded universal quantum computation  \cite{Jo09}. In classical computer science, finally, matchgates occur in various studies related to e.g. the theory of holographic algorithms \cite{Va02, Ca06}.

The aim of the present paper is to characterize the computational power of matchgate circuits. We will in particular study which  boolean functions\footnote{Henceforth, whenever we refer to a `function' we will often mean a boolean function. This will be clear from the context.} can be computed with such circuits. Given an arbitrary matchgate circuit $U$, the question is asked which boolean function $f(x)$ can be computed (probabilistically) by initializing the system in the computational basis state $|x, 0\rangle$ (where $x$ represents an input string and $0$ is a string of ancillary zeros), by subsequently running the circuit $U$ and finally measuring, say, the first qubit. This setting captures in perhaps the most elementary way the computational power of matchgate circuits, associating which each  circuit a yes/no question as commonly done.

We remark that, beyond its natural computer scientific interest, such an investigation is relevant from a intrinsic physical perspective as well. In particular, given the aforementioned equivalences between matchgate circuits, fermionic systems and 1D spin systems, the present work aims at gaining insight in the link between the physics of these systems and their computational capabilities. 
In this context one may pose a variety of interesting questions such as: `Does the presence of a quantum phase transition in the XY model leave any signature in the associated class of functions which can be computed by time-evolving such systems?' The present work is also situated within such a program.

In the following we will characterize the family of matchgate-computable functions in full generality. We will find that this class precisely coincides with the class of \emph{linear threshold gates} (LTGs) \cite{Der65}. The latter is a fundamental family of functions that has been a topic of study since the 1960s and that plays an important role in numerous areas.  LTGs occur in the study of neural networks where these functions serve as elementary models of neurons \cite{Mi88} and in circuit complexity theory \cite{Go92}; cf. also e.g. \cite{Ve06} and references within for a number of recent investigations on LTGs.
The existence of a connection between matchgates and LTGs may be considered surprising, since a priori there is no obvious relation between these two theories.

Below we state the contributions of this work in more precise terms. Here we highlight some particularly noteworthy aspects of our results.

An interesting phenomenon occurs when considering those functions that are matchgate-computable with \emph{high} success probability. Since a generic matchgate circuit may be  a rather complex object, one would expect that this class of functions has some nontrivial character as well.  We will show that this intuition is incorrect: any function that is matchgate-computable with success probability greater than $3/4$ is proved to be \emph{trivial} in that any such  function can only depend on a \emph{single} bit of the input.
The origin of this apparent paradox is the strong set of constraints that are placed on any circuit (matchgate or conventional) that is to compute the correct function value with high probability on \emph{all} inputs. Indeed, a generic circuit does typically \emph{not} fulfill these requirements and hence does not meaningfully compute any function. The present example thus highlights the significance of bounded-error constraints in a rather striking way: in spite of the potentially complex structure of general matchgate circuits,  the only instances which turn out to satisfy the bounded-error constraints are circuits  computing mere single-bit---i.e. utterly trivial---functions!

Note that the above feature is of interest from a physics perspective as well: it shows that the nontrivial \emph{physical} properties (such as e.g. the presence of a quantum phase transition)  of a family of quantum systems are \emph{not} guaranteed to translate into any nontrivial associated \emph{computational} model. Indeed, in spite of the interesting physical processes modeled by matchgates, these processes turn out to have trivial power when used as a computer which is to solve problems with high success probability.

Overall, the results in this paper indicate that matchgate circuits have a rather weak computational power. This is in part reflected in the above phenomenon, but also in an additional result obtained in this work. We will characterizes the power of matchgate circuits in terms of a strikingly elementary \emph{classical} computer which is capable of computing every matchgate-computable function with the same success probability as the optimal matchgate circuit.

\

{\bf Notation.---} In the following $[n]$ denotes the set of positive integers from 1 to $n$, for any $n$. If $x=(x_1, \dots, x_n)$ is a string of bits $x_k\in\{0, 1\}$, then $\hat x$ denotes the $\pm 1$ vector that is obtained by replacing all 0-components of the $n$-bit string $x$ by $1$ and all $1$-components by $-1$. The symbol $\oplus$ denotes addition modulo 2. The 1-norm of a complex vector $w=(w_1, \dots, w_n)$ is denoted by $\|w\|_1:=\sum_k |w_k|$.

\section{Main results}\label{sect_statement}

Here we summarize the contributions of this paper. To do so, we state some preliminary definitions.

A matchgate\footnote{The term `matchgate' sometimes refers to a larger class of operations (tensors) which may be both non-unitary and which may act on more than two qubits,  containing the unitary two-qubit gates (\ref{matchgate}) as a subclass; cf. \cite{Ca06, Br09}. In this paper, however, a matchgate is always taken to be a unitary two-qubit operation as defined in (\ref{matchgate}).} $G$ is any two-qubit operation with matrix representation \be\label{matchgate} G = \left[\begin{array}{cccc} a & & & b \\ & u&v&\\&x&y& \\ c & & & d\end{array} \right],\quad A =\left[\begin{array}{cc} a & b\\ c & d\end{array} \right], \quad B=\left[\begin{array}{cc} u & v\\ x & y\end{array} \right] \ee in the standard basis, where $A$ and $B$ belong to $SU(2)$.  We will consider circuits composed of matchgates acting on nearest-neighbor qubits (assuming a one-dimensional ordering of the qubits) i.e. the standard scenario in which matchgates are considered.

Let $f:\{0, 1\}^n\to\{0, 1\}$ be a boolean function on $n$ bits and let $U$ be an $m$-qubit unitary operation with $m\geq n$. We say that $U$ computes $f$ with probability at least $p$ if for every $n$-bit string $x$, the preparation of the $m$-qubit state $U|x, 0\rangle$ (where $0$ denotes a string of $m-n$ zeroes) followed by a computational basis measurement of the first qubit yields the outcome $f(x)$ with probability at least $p$. An $n$-bit boolean function $f$ is said to be matchgate-computable with probability at least $p$  if there exists a matchgate circuit on $m$ qubits, for some $m\geq n$, which computes $f$ with probability at least $p$. Note that in the latter definition no restriction is placed on the size of $m$ as compared to $n$, nor on the number of gates in the matchgate circuit compared to $n$. However, below we will find that every matchgate-computable function can be computed by a matchgate circuit acting on at most $n+1$ qubits (without decreasing the success probability). Moreover, it is known that any matchgate circuit acting on $m$ qubits can be re-rexpressed as a matchgate circuit of size $O(m^3)$ (see  \cite{Jo08}).

\

{\bf I. Main Theorem.---} We will show that the class of matchgate-computable functions coincides with the family of linear threshold gates. A boolean function  $f$ on $n$ bits is called a linear threshold gate (LTG) if there exist and an $n$-dimensional real vector $w$ and a real constant $\theta$ such that $f(x)$ equals 0 if and only if $w^T\hat x +\theta$ is strictly positive.  The vector $(w, \theta)$ is called a representation of $f$. Examples of linear threshold gates are the NOT gate, the $n$-bit AND and OR and the majority gate.
We will consider linear threshold gates that are supplemented with a parameter which is called the margin of the LTG. Given a linear threshold gate $f$ on $n$ bits with representation $(w, \theta)$, the margin of this representation is defined to be the minimal value of $|w^T\hat x +\theta|$ over all $n$-bit strings $x$. The margin $\epsilon$ of $f$ itself is the maximal achievable margin of any representation $(w, \theta)$ of $f$ which is normalized in the sense that $\|w\|_1 + |\theta| = 1$.

The main result of this paper achieves a complete characterization of all matchgate-computable functions:

\begin{thm}\label{thm_gaussian}
Let $f$ be a boolean function on $n$ bits and let $p\in (0.5, 1]$. Then  the following statements are equivalent:
\begin{itemize}
\item[(a)]  $f$ is matchgate-computable with probability at least $p$.
\item[(b)] $f$ is a linear threshold gate with margin $\epsilon\geq 2p-1$.
\end{itemize}
Moreover, (a) holds if and only if there exists a matchgate circuit acting on at most $n+1$ qubits which computes $f$ with probability at least $p$.
\end{thm}
\noindent Note that theorem \ref{thm_gaussian} connects the margin $\epsilon$ of an LTG with the optimal success probability $p$ of computing this function using matchgate circuits. In particular,  $\epsilon$ is small iff  $p$ is small. This implies that the class of linear threshold gates computable with matchgate circuits grows larger as the required probability of success is decreased. When $p$ is allowed to be arbitrary close to $0.5$, the full class of LTGs is matchgate-computable due to theorem \ref{thm_gaussian}.

As an immediate corollary of the above result, it  follows that matchgate circuits do not have universal classical computational power, even when an unbounded error is allowed i.e. a success probability strictly greater then $0.5$ but (with increasing  $n$) potentially exponentially close to 0.5. This property follows immediately from the elementary fact that there exist functions that are not LTGs, such as the two-bit parity gate.

\

{\bf II. Bounded-error computations.---} Surprisingly, it follows from theorem \ref{thm_gaussian} that matchgate circuits can only compute trivial functions if the computation is to succeed with high probability:
\begin{thm}\label{thm_large_p}
A boolean function $f$ is matchgate-computable with probability $p>\frac{3}{4}$ if and only if this function is either constant or  depends on a single bit of its input.
\end{thm}
\noindent  Due to theorem \ref{thm_gaussian}, any function that is matchgate-computable with probability $p>3/4$ is a linear threshold gate with margin $\epsilon>1/2$. We will show that the only LTGs having such large margin are constant or  depend on one input bit, leading to the proof of theorem \ref{thm_large_p}. Note that functions which depend on a single input bit have the very simple form $f(x) = x_k$ or $f(x)=1-x_k$ for some $k$, i.e. $f$ returns the $k$-th bit of its input or its negation. As discussed in the introduction, theorem \ref{thm_large_p} is a somewhat unexpected result, given that generic matchgate circuits are rather nontrivial objects which e.g. describe physical systems that may exhibit interesting behavior such as e.g. quantum phase transitions. In spite of this rich structure, theorem \ref{thm_large_p} shows that the computational power of matchgate circuits is near-vanishing in those cases where the correct answer is to be produced with high probability.

Of particular interest are computations  which yield the correct output with a probability that is bounded away from $0.5$ by an inverse polynomial in the input size. Let ${\cal F}=\{f_n: n=1, 2, \dots\}$ be a family of boolean functions where $f_n$ acts on $n$ bits and let $\{p_n\}$ be a family of probabilities where $p_n>0.5$ and, for $n$ large, $p_n$ is bounded away from $0.5$ by an inverse polynomial  in $n$. We say that ${\cal F}$ is matchgate-computable with  poly-bounded error if there exists such a poly-bounded family of probabilities as well as a family of matchgate circuits $\{U_n\}$, such that $U_n$ computes $f_n$ with probability at least $p_n$. It is standard that any computation with poly-bounded error can be promoted to an almost-deterministic computation\footnote{That is, the success probability is exponentially (in $n$) close to 1.} by repeating the computation poly$(n)$ times and taking the majority vote of all obtained results. Due to theorem \ref{thm_gaussian}, the families of boolean functions that are matchgate-computable with poly-bounded error precisely coincides with those LTGs having poly-bounded  margin i.e. the margin $\epsilon_n$ of $f_n$  scales an an inverse polynomial in $n$. We will furthermore show that a family of LTGs has poly-bounded margin if and only if $f_n$ has a representation $(w_n, \theta_n)$ where the coefficients of $w_n$ and $\theta_n$ are \emph{integers} that are at most polynomially  large; such families of threshold gates are said to have polynomial  integer weight. This leads to the following concise characterization.
\begin{thm}\label{thm_poly_bounded}
A family of boolean functions is matchgate-computable with poly-bounded error if and only if it is a family of linear threshold gates with polynomial integer weight.
\end{thm}
{\it Remark.---}  The near-determinstic computation of $n$-bit LTGs with polynomial integer weight is obtained by running the computation poly$(n)$ times and computing the majority vote of all measurement outcomes. It is intriguing that the majority function is indeed an LTG---and hence matchgate-computable---however in order to properly amplify the success probability, the majority gate itself needs to be computed with suitably high success probability (the latter e.g. being  exponentially close to 1). However, due to theorem \ref{thm_large_p} the majority function \emph{cannot} be computed with good success probability by any matchgate circuit. Therefore, to obtain the proper amplification a final \emph{non}-matchgate computation is needed to compute the majority vote---even though the latter comes intriguingly close to being suitably matchgate-computable! \hfill $\diamond$

\

{\bf III. An equivalent classical computer.---} As a final result, we will construct a \emph{classical} computational scheme that is equivalent to the matchgate circuit model in the following sense: any function which is matchgate-computable with probability at least $p$ can be computed with probability at least $p$ with our classical scheme, and vice versa.  As we will show, the required classical computer is very simple, as it essentially requires a single sample of a fixed (i.e. independent of the input) probability distribution on the set of integers from 1 to $n+1$,  together with the possibility of performing a single bit flip of one of the input bits, depending on the outcome of the sampling. This characterization is a further illustration (in addition to e.g. theorem \ref{thm_large_p}) of the weak computational capabilities of matchgate circuits.

Whereas the general result will be stated below, here we illustrate the scheme with an example. Let $n$ be odd and consider the majority function $f_{\mbox{\scriptsize maj}}$ on $n$-bits, which is an LTG. It can be shown that the margin of $f_{\mbox{\scriptsize maj}}$ is $\epsilon_{\mbox{\scriptsize{maj}}}=n^{-1}$. Due to theorem \ref{thm_gaussian}, the optimal success probability of computing  $f_{\mbox{\scriptsize maj}}$  with a matchgate circuit is \be\label{prob_maj} p_{\mbox{\scriptsize{maj}}} = \frac{\epsilon_{\mbox{\scriptsize{maj}}} +1}{2} = \frac{1}{2} + \frac{1}{2n}.\ee  Now consider the following elementary classical computation:
\begin{itemize}
\item Choose an $n$-bit input string $x$;  \item Generate a random integer  $k$ between 1 and $n$;
 \item Output the bit value $x_k$.
\end{itemize}
It can easily be shown that, for every $x$, the output of this computation is $f_{\mbox{\scriptsize{maj}}}(x)$ with probability at least $p_{\mbox{\scriptsize{maj}}}$. In other words, the above elementary classical computation is capable of computing the majority function with the same probability of success as the optimal matchgate circuit can!

The above example is not  coincidental, as we will show that \emph{every} LTG $f$ can similarly be associated with a simple classical computation of the above nature.

\section{Matchgates}\label{sect_matchgates}

In this section we recall some basic properties of matchgates, which were defined in Eq. (\ref{matchgate}). We emphasize that, henceforth, the term `matchgate circuit' will always refer to a circuit composed of matchgates acting on nearest-neighbor qubit lines, as commonly done.

We first recall a celebrated result about the classical simulation of matchgate circuits, first proved in \cite{Va02} and subsequently investigated by a series of other authors \cite{Di04, Ca06, Jo08, Br09}.

\begin{thm}
Consider a poly-size $n$-qubit circuit composed of matchgates acting on nearest-neighbor qubits. The circuit acts on an arbitrary standard basis input and is followed by a  standard basis measurement of the first qubit. Let $p_0$ denote the probability of obtaining the bit 0 as an outcome. Then there exists a classical algorithm which computes $p_0$ up to $m$ bits in poly$(n, m)$ time.
\end{thm}
\noindent As a consequence of this result, there exists a poly-time classical algorithm to sample from (a distribution that is exponentially close to) the distribution $\{p_0, 1-p_0\}$ i.e. any matchgate computation of the above type can be simulated classically in poly-time.

We discuss some further well-established  properties of matchgates, which will be used in the proof of theorem \ref{thm_gaussian}. We refer to e.g. \cite{Jo08} for elementary proofs of these properties.

First, without loss of generality, in the following we will always consider poly-size families of matchgate circuits, as it is known that any, possibly exponential size, $n$-qubit matchgate circuit family can be re-expressed a matchgate circuit family of size $O(n^3)$.  Second, consider the $n$-qubit Jordan-wigner operators:
\be\label{JordanWigner} c_{2k-1} = X_k\prod_{i=1}^{k-1} Z_i,\quad c_{2k} = Y_k\prod_{i=1}^{k-1} Z_i,\ee  where $X_k$, $Y_k$ and $Z_k$ denote the Pauli $X$, $Y$ and $Z$ operators acting on qubit $k$, respectively. Then an $n$-qubit unitary operation $U$  is a matchgate circuit (up to a global phase) if and only if there exists an operator $R\in SO(2n)$ such that, for every $\mu\in[2n]$, it is the case that \be\label{SO2n} U^{\dagger} c_{\mu} U = \sum_{\nu=1}^{2n} R_{\mu\nu} c_\nu.\ee
It further holds that $U$ is a matchgate circuit if and only if there exists a hermitian operator $H$ lying in the linear span of the products $c_{\mu}c_{\nu}$ (where $\mu\neq\nu$), such that $U\propto e^{i H}$. Such an operator $H$ is sometimes called a quadratic hamiltonian. Quadratic hamiltonians describe the physics of systems of non-interacting fermions. We will not discuss this connection to fermionic physics here as it would lead us too far outside of the scope of this work, and we refer to e.g. \cite{Di04}.

The following are some examples of matchgates and matchgate circuits. The fermionic SWAP  (fSWAP) operation
is easily seen to be a matchgate. This operation sends the basis state $|ab\rangle$ to $(-1)^{ab} |ba\rangle$, for every $a, b=0, 1$, i.e. it swaps the qubits and adds an overall minus sign if both qubits are in the state $|1\rangle$. Other elementary examples of matchgate circuits are the products $c_{\mu}c_{\nu} $ for any $\mu\neq\nu$. Indeed, any such product can be written as an exponential of a quadratic hamiltonian. Denoting $H:= ic_{\mu}c_{\nu}$ it is easily verified that $H=H^{\dagger} = H^2$. Hence \be e^{i\frac{\pi}{2} H} \propto H  \propto c_{\mu}c_{\nu},\ee where  in the first identity we have used that $e^{itH} = (\cos t) I + i(\sin t) H$ since $H=H^2$. To obtain a more nontrivial example of a matchgate circuit, consider the hamiltonian of the one-dimensional transverse ising model: \be H_I = -\sum_{k=1}^{n-1} J_k X_k X_{k+1} - \sum_{k=1}^n h_k Z_k,\ee  for some real constants $J_k$ and $h_k$. It can readily be verified that \be X_k X_{k+1} \propto  c_{2k}c_{2(k+1)-1} \quad\mbox{ and }\quad Z_k \propto c_{2k-1}c_{2k},\ee showing that $H_I$ is a quadratic Hamiltonian. Therefore, for every real $t$ the time evolution operator $e^{itH_I}$ can be written as a (poly-size) matchgate circuit.

The notion  of a matchgate-computable boolean function was introduced in section \ref{sect_statement}. Here we discuss some simple examples. It can easily be seen that every boolean function which is either constant or which depends on a single input bit can computed with unit probability by an elementary matchgate circuit composed of operators $c_{\mu}c_{\nu}$ and fSWAP gates, acting on $n+1$ qubits. Note that the only possible functions of this type are the functions $x\to 0$,  $x\to 1$,  $x\to x_k$ and $x\to 1-x_k$, for some $k$. We show that $x\to 1-x_k$ is matchgate-computable with unit probability; the other three cases are treated similarly. For each $k\in[n]$, denote the operator $U_k$ acting on $n+1$ qubits by \be U_k:=c_{2k-1}c_{2(n+1)-1}\propto X_k X_{n+1} \prod_{j=k}^{n} Z_j.\ee Note  that $U_k|x, 0\rangle\propto| x_{\neg k}, 1\rangle,$ where $x_{\neg k}$ denotes the bit string obtained by flipping the $k$-th bit of $x$. Now consider an elementary matchgate computation where first the state $U_k|x, 0\rangle$ is prepared,  followed by measurement of qubit $k$. This computation yields the bit $1-x_k$ with unit probability.  This shows that the desired function can be computed with unit probability by applying a suitable matchgate circuit and measuring \emph{some}  qubit in the computational basis. By applying a suitable sequence of fSWAP gates immediately before the measurement, we may assume w.l.o.g. that the first qubit is measured. Indeed, as the measurement is in the computational basis, the minus sign of the fSWAP gate has no relevance, and this gate acts as a simple SWAP.

Remarkably, as we will prove in theorem \ref{thm_large_p}, constant and single-bit functions are the \emph{only} functions that are matchgate-computable with high success probability.

\section{Linear threshold gates}\label{sect_LTGs}

In this section we discuss some elementary features of linear threshold gates (LTGs).
For convenience we recall here their definition. An $n$-bit boolean function $f$ is a linear threshold gate if there exists an $n$-dimensional real vector $w$ and a real number $\theta$ such that \be\label{def_LTG} (-1)^{f(x)} = \mbox{sign}(w^T\hat x + \theta)\ee for every $n$-bit string $x$. The pair $(w, \theta)$ is called a representation of $f$.  Definition (\ref{def_LTG}) has an elementary geometrical interpretation. Taking an arbitrary $w$ and $\theta$, the linear equation $w^T z+\theta=0$ defines a hyperplane which divides the $n$-dimensional real space in two parts, say $H_{+}$ and $H_{-}$, where $H_{+}$ consists of all $z\in\mathbb{R}^n$ such that $w^Tz+\theta\geq 0$ and $H_{-}$ is defined as the complement of $H_{+}$. The LTG $f$ associated with $(w, \theta)$ then simply evaluates whether a given $\{\pm 1\}$-vector $\hat x$ lies in $H_{+}$ or $H_{-}$. Stated differently, a boolean function $f$ is a linear threshold gate iff there exists a hyperplane in $n$-dimensional real space which separates the sets of inputs $x$ that are mapped to 0 and 1, respectively.

It can easily be verified that the constant and single-bit functions are linear threshold gates. Other examples are the the $n$-bit AND, OR and majority function. The AND gate, for example, has a representation $(w, \theta)$ given by $w=(1, \dots, 1)$ and $\theta= -n+\frac{1}{2}$. It is also known that not all functions are LTGs. For example, the two-bit parity gate \be f(x_1, x_2):=x_1\oplus x_2\ee is  not a linear threshold gate, as can be easily verified.

Every linear threshold gate has infinitely many representations. For example, rescaling the vector $(w, \theta)$ with a positive multiplicative constant trivially leads to the same associated function. We will say that the representation $(w, \theta)$ is normalized if the 1-norm of this vector is equal to 1, i.e. $\|w\|_1 + |\theta| =1$. Interestingly, every linear threshold gate has a representation $(w, \theta)$ where each $w_k$ and $\theta$ are \emph{integers}. The intuition behind this result is the following. It is easy to show that every LTG has a representation $(w, \theta)$ where each $w_k$ and $\theta$ are rational numbers, say $w_k = \frac{a_k}{b_k}$ and $\theta = \frac{c}{d}$ for suitable integers $a_k, b_k, c$ and $d$; this essentially follows from the fact that the rationals are dense in the reals and the property that small perturbations of a representation do not change the associated LTG. Multiplying the rational representation $(w, \theta)$ with the product $\{\prod |b_k|\}|d|$ then yields an integer representation.

Further, it is known that every linear threshold gate on $n$ bits has a representation $(w, \theta)$ where each of the components of $w$, as well as the number $\theta$, are integers not greater than $2^{N}$ in absolute value, with $N=O(n\log n)$ \cite{Mu61}; moreover there exist linear threshold gates where such large numbers are required \cite{Ha94}. This shows that every linear threshold gate $f$ admits a representation that can be fully described in terms of $O(n^2\log n)$ bits, and that $f(x)$ can be evaluated in poly-time when this particular representation is provided.

Finally, we recall that that, given a representation  $(w, \theta)$ of an LTG $f$, the margin of this representation is defined to be the minimal value of $|w^T\hat x +\theta|$ over all $n$-bit strings $x$. The margin $\epsilon(f)\equiv \epsilon$ of $f$ is then the maximal margin over all \emph{normalized} representations. Note that $\epsilon(f)$, which should in principle be defined as a supremum, is indeed a maximum. This can be argued with standard methods\footnote{To see this, first note that the margin of a representation $(w, \theta)\equiv \mathbf{z}$ is a continuous function of $\mathbf{z}$, being defined as the minimum of a finite number (i.e. $2^n$) of continuous functions $g_x(\mathbf{z}):=|w^T\hat x+\theta|$ for every $x$. Thus $\epsilon(f)$ is the supremum of a continuous function in $z$, over all representations $\mathbf{z}$ of $f$ with $\|\mathbf{z}\|_1=1$. As the set of all such normalized representations is a compact set, it follows that the supremum is reached.}. In section \ref{sect_margins} we will focus in more detail on the properties of margins of linear threshold gates, which will lead to the proofs of theorems \ref{thm_large_p} and \ref{thm_poly_bounded}. Before doing so, we provide the proof of theorem \ref{thm_gaussian} in the next section.

\section{Proof of theorem \ref{thm_gaussian}}

In this section we prove theorem \ref{thm_gaussian}.  The proof will proceed in three steps. In step 1 we  reduce the study of matchgate-computable functions to the investigation of matrix elements of the form $\langle x|U^{\dagger}Z_1U|x\rangle$, where $|x\rangle$ denotes a computational basis state, $U$ is a matchgate circuit and $Z_1$ is the Pauli $Z$ operator acting on the first qubit. In step 2, which represents the main ingredient of the proof of theorem \ref{thm_gaussian}, the most general form of such matrix elements is characterized. Finally, in step 3 the proof is completed by combining steps 1 and 2.

\subsection{Step 1}\label{sect_step1}

Let $f$ be an $n$-bit boolean function, let $U$ be an arbitrary $m$-qubit unitary operation for some $m\geq n$ and fix $p\in(0.5, 1]$.  Furthermore, we denote $\langle Z\rangle_x:= \langle x, 0|U^{\dagger}Z_1U|x, 0\rangle$. We now state the following claim.

\

\noindent {\bf Claim.} {\it $U$ computes $f$ with probability at least $p$ if and only if, for every $x$, one has   \begin{itemize}
\item[(a)] $|\langle Z\rangle_x| \geq 2p-1$ and \item[(b)] $\mbox{sign}\langle Z\rangle_x = (-1)^{f(x)}$.\end{itemize}
}

\noindent To prove this claim, consider the preparation of $U|x, 0\rangle$ followed by a computational basis measurement of the first qubit and let $p_x$ denote the probability that the measurement outcome is $f(x)$, for every $x$. Then $U$ computes $f$ with probability at least $p$ if and only if $p_x\geq p$ for every $x$. We thus have to show that (a)-(b) are equivalent to the condition $p_x\geq p$ for all $x$.

We will distinguish between the cases $f(x)=0$ and $f(x)=1$. We start with the former case.
As $p_x$ and $1-p_x$ are the probabilities of obtaining the measurement outcomes 0 and 1, resp.,  one has \be\label{eq_conditions_ab} \langle Z\rangle_x = p_x - (1-p_x) = 2p_x -1.\ee
Now suppose first that $p_x \geq p$ for every $x$ (with $p\in(0.5, 1]$ as stated above). Conditions (a) and (b) then follow immediately. Conversely, assume that (a) and (b) are true.  As (b) is satisfied, we have $ \langle Z_1\rangle_x\geq 0$ and using (a) it thus follows that $ \langle Z_1\rangle_x\geq 2p-1$. Invoking (\ref{eq_conditions_ab}) then implies that  $p_x\geq p$ for every $x$, as desired. This completes the proof for the case $f(x)=0$.

The case $f(x)=1$ is treated analogously; the main distinction is that now $p_x$ represents the probability of measuring 1. Consequently, (\ref{eq_conditions_ab}) is replaced by \be\langle Z\rangle_x = (1-p_x) - p_x = 1- 2p_x.\ee The remainder of the argument is completely analogous.

\subsection{Step 2}
Conditions (a)-(b) imply that the study of matchgate-computable functions reduces to the investigation of matrix elements of the form $\langle x, 0|U^{\dagger}Z_1U|x, 0\rangle$, where $U$ is an arbitrary matchgate circuit. Next we investigate the most general form which such matrix elements may take. A complete characterization of this problem is obtained in the following theorem.

\begin{thm}\label{thm_diag}
Let $U$ be an $n$-qubit unitary operator. If $U$ is a matchgate circuit then there exists an $a\in\mathbb{R}^n$ with $\| a\|_1\leq 1$ such that \be\label{eq_lem_diag} \langle x|U^{\dagger} Z_1 U|x\rangle = a^T\hat x \ee for every $n$-bit string $x$. Conversely, for every $a\in\mathbb{R}^n$ with $\| a\|_1\leq 1$  there exists an $n$-qubit matchgate circuit $U$ such that
(\ref{eq_lem_diag}) holds.
\end{thm}

In the proof of this theorem we will need the following elementary fact.

\begin{lem}\label{thm_one-norm}
Consider  a  vector $a\in\mathbb{R}^n$ with $\|a\|_1\leq 1$. Then there exist $u, v\in\mathbb{R}^n$  with $\| u\|_2= 1$ and $\| v\|_2 \leq 1$ such that $a_k:= u_kv_k$ for every $k\in[n]$.
\end{lem}
\noindent {\it Proof:} Define $u$ and $v$ by \be u_k:= \sqrt{|a_k|/\|a\|_1} \quad\mbox{ and }\quad v_k:= \sqrt{|a_k|\cdot \|a\|_1}\cdot  \mbox{ sign}(a_k),\ee for every $k\in[n]$, respectively. Obviously, $a_k = u_k v_k$. Moreover, $\| u\|_2= 1$ and $\| v\|_2 = \|a \|_1\leq 1$, as can be easily verified. This proves the lemma. \finpr

\

{\it Proof of theorem \ref{thm_diag}: } We first prove the forward direction. Denote $O:= U^{\dagger}Z_1U$ and let $R$ be the $SO(2n)$ rotation associated to $U$ via (\ref{SO2n}). Letting $\rho$ and $\rho'$ denote the first, resp. second, row of $R$ and using that $Z_1 = -i c_1c_{2}$, it follows that \be O = -i[U^{\dagger}c_1U][ U^{\dagger} c_2 U] = \sum \rho_{\mu} \rho'_{\nu} [-ic_{\mu}c_{\nu}],\ee where the sum is over all $\mu, \nu\in[2n]$. As $\langle x|O|x\rangle$ is a diagonal entry of $O$ for every $x$, we only need to focus on the diagonal part of this operator i.e. $\mbox{diag}(O):=\sum_x \langle x|O|x\rangle |x\rangle\langle x|$. Using the explicit representation (\ref{JordanWigner}) of the $c_{\mu}$, it is easily verified that
\be \mbox{diag}(-ic_{\mu}c_{\nu})=\left\{
\begin{array}{cl} -iI & \mbox{ if } \mu = \nu\\ Z_k &\mbox{ if } (\mu, \nu)= (2k-1, 2k) \mbox{ for
some } k\in[n]
\\ -Z_k &\mbox{ if } (\mu, \nu)= (2k, 2k-1) \mbox{ for some } k\in[n]
\\ 0 & \mbox{ otherwise.}\end{array} \right.\ee
Setting $a_k:= \rho_{2k-1} \rho'_{2k} - \rho_{2k} \rho'_{2k-1}$ for every $k\in [n]$ and using that $\sum_{\mu}\rho_{\mu} \rho'_{\mu}$ is zero since $\rho$ and $\rho'$ are two distinct rows of an orthogonal matrix, it follows that \be \mbox{diag} (O) = \sum_{k=1}^n a_k Z_k.\ee
The $(x, x)$ diagonal entry of $O$ thus reads: \be \langle x|O|x\rangle = \sum a_k \langle x|Z_k|x\rangle = a^T\hat x.\ee   This shows that that (\ref{eq_lem_diag}) is satisfied. Note also that $|\langle x|O|x\rangle|\leq 1$ for every $x$ since $O$ is unitary. Moreover,  it is easily verified that there exists an $x$ such that $a^T\hat x = \| a\|_1$. This shows that $\| a\|_1 \leq 1$.

To prove the reverse direction, consider an arbitrary $a\in\mathbb{R}^n$ with $\| a\|_1\leq 1$. Due to lemma \ref{thm_one-norm}, there exist $n$-dimensional real vectors $u$ and $v$ with $\| u\|_2 = 1$ and $\| v\|_2\leq 1$ such that $a_k = u_kv_k$ for every $k\in [n]$.  Furthermore, it is elementary that for every such $u$ and $v$ there exists a vector $w\in \mathbb{R}^n$ that is orthogonal to $u$ and that satisfies $ \|w\|_2^2 =1 - \| v\|_2^2 $. Consequently, the $2n$-dimensional vectors \be \rho :=(u_1, 0, u_2, 0, \dots, u_n, 0)\quad\mbox{ and }\quad\rho':= (w_1, v_1, w_2, v_2, \dots, w_n, v_n)\ee are unit vectors (w.r.t. the 2-norm) that are orthogonal. Let $R$ be any $SO(2n)$ rotation having $\rho$ and $\rho'$ as first, resp. second row. Let $U$ be the $n$-qubit  matchgate circuit associated to $R$. Analogous to the proof of the forward direction of the theorem, a direct calculation shows that the diagonal part of $U^{\dagger} Z_1 U$ equals \be \sum [\rho_{2k-1} \rho'_{2k} - \rho_{2k} \rho'_{2k-1}]Z_k = \sum_{k=1}^n a_k Z_k.\ee It immediately follows that (\ref{eq_lem_diag}) is satisfied. This completes the proof.
\finpr

\subsection{Step 3}
We now show how the above result leads to the proof of  Theorem  \ref{thm_gaussian}.  Invoking Step 1, it suffices to show that, for every $n$-bit boolean function $f$ and $p\in(0.5, 1]$, the following are equivalent:
\begin{itemize}
\item There exists an $m$-qubit matchgate circuit $U$, for some $m\geq n$, such that conditions (a) and (b) in Step 1 hold;
\item $f$ is a linear threshold gate with margin $\epsilon\geq 2p-1$.
\end{itemize}
We first prove the forward direction of the claim.  Let $U$ be a matchgate circuit on $m$ qubits such that (a)-(b) hold.
We first invoke theorem \ref{thm_diag}. This yields a vector $a\in\mathbb{R}^{m}$ with 1-norm at most 1 such that (\ref{eq_lem_diag}) holds. With the notations $\bar a:=(a_1, \dots, a_n)$ and $b:= \sum_{l=n+1}^{m} a_l$ it then immediately follows that \be\label{eq_thm_1} \langle x, 0| U^{\dagger}Z_1U |x, 0\rangle ={\bar a}^T\hat x + b\ee for every $n$-bit string $x$. Together with condition (b) in Step 1 this shows that $f$ is a linear threshold gate with representation $(\bar a , b)$. Normalizing the vector $(\bar a, b)$ w.r.t. the 1-norm yields a normalized representation $w:= \gamma \bar a$ and $\theta := \gamma b$, where $\gamma^{-1}:= \|\bar a\|_1 + |b|.$ Note that $\gamma^{-1}\leq \|a\|_1\leq 1$. Using Eq. (\ref{eq_thm_1}) and condition (a) of Step 1 it follows that \be |{\bar a}^T\hat x + b|\geq 2p-1.\ee Moreover, using the definitions of $w$ and $\theta$ and the fact that $\gamma\geq 1$, it finally follows that \be |w^T\hat x + \theta| \geq\gamma (2p-1)\geq 2p-1,\ee for every $x$. This shows that the margin of $f$ is at least $2p-1$.

We now show the converse.  Consider an $n$-bit LTG $f$.  Let $\epsilon\in(0, 1]$ be its margin  with associated (normalized) representation $(w, \theta)$ and denote $p:=(\epsilon + 1)/2$. Due to theorem \ref{thm_diag}, there exists an $(n+1)$-qubit  matchgate circuit $U$ such that \be \langle x, y| U^{\dagger}Z_1U |x, y\rangle = w^T \hat x + \theta \hat y\ee for every $n$-bit string $x$ and for every $y=0, 1$. This readily implies that \be \mbox{sign}(\langle x, 0| U^{\dagger}Z_1U |x, 0\rangle) = \mbox{ sign}(w^T\hat x + \theta)= (-1)^{f(x)},\ee for every $x$.  Moreover, as $\epsilon$ is the margin of $f$, it follows that \be |\langle x, 0| U^{\dagger}Z_1U |x, 0\rangle| = |w^T \hat x + \theta|\geq 2p -1\ee for every $x$.This shows that conditions (a)-(b) in Step 1 are fulfilled. This completes the proof of theorem \ref{thm_gaussian}.

{\it Remark.---} It follows from the above argument that an $n$-bit boolean function $f$ is computable by an $m$-qubit matchgate circuit with probability at least $p$, for \emph{some} $m\geq n$, if and only if  $f$ is computable with probability at least $p$ by a matchgate circuit acting on at most $n+1$ qubits.
\hfill $\diamond$

\section{Margins of threshold gates}\label{sect_margins}

In this section we analyze the properties of margins of LTGs in more detail. In particular, the proofs of  theorems \ref{thm_large_p} and \ref{thm_poly_bounded} will follow from the considerations in this section.

\subsection{Large margins}

Next we investigate the subclass of linear threshold gates with large margin. We will in particular show that any LTG with  margin strictly larger than $1/2$ is essentially trivial. The proof of theorem \ref{thm_large_p} will follow immediately from this property.

Recall that, formally, a boolean function $f$ on $n$ bits is said to depend on its $k$-th variable, if there exists an $n$-bit string $x$ such that $f(x) \neq f(x_{\neg k})$, where $x_{\neg k}$ is the string obtained by flipping the $k$-th bit of $x$. We can now state the following result:

\begin{lem}\label{thm_LTG_variables}
Any linear threshold gate with margin $\epsilon$ can depend on at most $t=1/\epsilon$ of its variables.
\end{lem}
{\it Proof:} Let $f$ be an $n$-bit LTG and let $\epsilon$ be the margin of $f$ with associated normalized representation $(w, \theta)$. We make the following claim. {\it Claim:} If $f$ depends on its $k$-th variable then $|w_k|\geq \epsilon$.
The proof of the lemma immediately follows from correctness of this claim, by using that $\|w\|_1\leq 1$ as $(w, \theta)$ is a normalized representation. We now prove the claim. For simplicity but without loss of generality, we set $k:=1$ and assume that $f$ depends on its first variable. We prove that $|w_1|\geq \epsilon$. Denote $\bar w:=(w_2, \dots, w_n)$. If $f$ depends on its first variable, there must exist an $\bar s\in\{\pm 1\}^{n-1}$ such that ${\bar w}^T\bar s \pm |w_1| + \theta$ have opposite signs. We consider the following two possibilities: (a) ${\bar w}^T\bar s + \theta\geq 0$ or (b)  ${\bar w}^T\bar s + \theta\leq 0$. In case (a), it follows that $ {\bar w}^T\bar s +|w_1| + \theta\geq 0$ and hence we must have ${\bar w}^T\bar s -|w_1| + \theta\leq 0$. Since $\epsilon$ is the margin of $(w, \theta)$, it follows that \be {\bar w}^T\bar s -|w_1| + \theta\leq -\epsilon.\ee 
Hence, $|w_1|\geq \epsilon$.  Case (b) is treated analogously.
\finpr

It follows that any linear threshold gate which depends on all its $n$ variables can admit a margin of size at most $1/n$, and any LTG with constant margin can depend on at most a constant number of its input bits.  What is more, if $\epsilon$ is sufficiently close to 1 viz. $\epsilon>1/2$ then $f$ can depend on at most one of its variables. Indeed, it follows from lemma \ref{thm_LTG_variables} that $f$ can depend on at most $t=\epsilon^{-1}<2$ input bits, i.e. $t$ is  0 (corresponding to a constant function) or 1. Combining these considerations with theorem \ref{thm_gaussian}, the forward direction of theorem \ref{thm_large_p} follows immediately: indeed, due to theorem \ref{thm_gaussian}  any boolean function which is matchgate-computable with probability $p>3/4$ must be an LTG with margin $\epsilon>1/2$, and hence of the form indicated. The proof of the converse direction of theorem \ref{thm_large_p} is elementary, as discussed in section \ref{sect_matchgates}.

More general than theorem \ref{thm_large_p}, we have actually showed that any function which can be computed by a matchgate circuit with success probability at least $p$ is an LTG which depends on at most  $O(p^{-1})$ of its input bits. As a consequence, whenever a constant success probability is considered (i.e. independent of the input size), only functions can be computed which depend on a constant number of input bits. Conversely, we will prove that any linear threshold gate depending on $k$ of its input bits can be computed by a matchgate circuit with success probability of at most $\frac{1}{2} + O(k^{-1})$. Therefore, any LTG that depends on all its $n$ input bits can be computed by a matchgate circuit with probability at most $\frac{1}{2} + O(n^{-1})$, which always lies at least polynomially close to 0.5.

We conclude this section with an example. Let $n$ be odd and let $f_{\mbox{\scriptsize{maj}}}$ denote the majority function on $n$ bits i.e. $f_{\mbox{\scriptsize{maj}}}(x)$ is 0
iff the $n$-bit string $x$ contains more zeros then ones. It can easily be verified that $f_{\mbox{\scriptsize{maj}}}$ is an LTG with normalized representation $(w, \theta)$  defined by $w_k:=n^{-1}$ for every $k$ and $\theta:=0$. Moreover, the minimal value of $|w^T\hat x + \theta|$ over all $x$ is easily shown  to be $n^{-1}$, showing that $n^{-1}$ is a lower bound for the margin of $f_{\mbox{\scriptsize{maj}}}$. As $f_{\mbox{\scriptsize{maj}}}$ depends on all its $n$ variables, its margin is at most $n^{-1}$ due to lemma \ref{thm_LTG_variables}. This shows that the margin of the majority function is precisely $n^{-1}\equiv \epsilon_{\mbox{\scriptsize{maj}}}$. Due to theorem \ref{thm_gaussian}, there hence exists a matchgate circuit that computes this function with probability \be\label{prob_maj} p_{\mbox{\scriptsize{maj}}} = \frac{\epsilon_{\mbox{\scriptsize{maj}}} +1}{2} = \frac{1}{2} + \frac{1}{2n}.\ee Moreover, any matchgate circuit computing the majority function can do so with probability at most $p_{\mbox{\scriptsize{maj}}}$.

\subsection{Margins and integer representations}

Next we focus on the possible types of asymptotic behavior  the margins of a family of LTGs may exhibit. We will in particular be interested in the distinction between margins that are polynomially bounded and margins that are exponentially small in the input size. As discussed in section \ref{sect_statement}, the subclass of (families of) LTGs with polynomially bounded margin coincides with those functions that are matchgate-computable with poly-bounded error. In the following we will in particular prove a simple characterization of the families of LTGs with poly-bounded margin.

An important parameter of an LTG $f$ in the present context will be the integer weight $\omega(f)\equiv \omega$ of $f$, defined to be the minimal 1-norm $\|w\|_1 + |\theta|$ of any possible integer representation of $f$. A family of LTGs $\{f_n: n=1, 2, \dots\}$, where $f_n$ acts on $n$ bits, is said to have polynomial integer weight if the integer weight $\omega_n$ of $f_n$ scales polynomially with $n$.
Interestingly, it turns out that the margin of an LTG and its integer weight are closely related concepts:

\begin{thm}\label{thm_margin_integer}
 Consider an $n$-bit linear threshold gate with margin $\epsilon$ and integer weight $\omega$. Then \be\label{margin_int_weight} \frac{1}{\epsilon} \leq \omega \leq  2\cdot \frac{n+1}{\epsilon}.\ee
\end{thm}
\noindent Note that, in the case of large $n$, the second inequality in (\ref{margin_int_weight}) simplifies to $\omega\leq O(\frac{n}{\epsilon})$.
As an immediate corollary of theorem \ref{thm_margin_integer}, the following property follows:
\begin{itemize}
\item[] {\it A family of linear threshold gates  has poly-bounded margin if and only if it has polynomial integer weight}.
\end{itemize}
Combining this result with theorem \ref{thm_gaussian} then proves theorem \ref{thm_poly_bounded}.

\

{\it Proof of theorem \ref{thm_margin_integer}:} We first prove the first inequality. Let  $(v, \varphi)$ be an integer representation with $\|v\|_1 + |\varphi| = \omega$. Normalizing $(v, \varphi)$ leads to the representation $(w, \theta)$ defined by $w:= \omega^{-1} v$ and $\theta = \omega^{-1}\varphi$. We claim that $(w, \theta)$ has margin at least $\omega^{-1}$, implying that $\epsilon\geq \omega^{-1}$ as desired. To prove the claim, note that the sign of $v^T \hat x + \theta$ equals $(-1)^{f(x)}$ for every $n$-bit string $x$. This implies that $v^T \hat x + \varphi\neq 0$ for any $x$. As the components of $v$ and $\varphi$ are integers, $v^T \hat x + \varphi$ is an integer as well for any $x$. Hence, the property  $v^T \hat x + \varphi\neq 0$ implies that $|v^T \hat x + \varphi|\geq 1$. It follows that $|w^T \hat x + \theta|\geq \omega^{-1}$ i.e. the margin of $(w, \theta)$ is at least $\omega^{-1}$.

Next we prove the second inequality. Let  $(w, \theta)$ be a normalized representation of $f$ with margin $\epsilon$.
Let $d$ be a positive integer; for now $d$ is arbitrary but we will fix a value later. Let $w_{k\mu}$ and $\theta_{\mu}$ represent the $\mu$-th bit in the binary expansion\footnote{Any real number $a\in[1, -1]$ can be expanded in a unique way as $a= \mbox{ sign}(a) \sum_{\mu=1}^{\infty} a_{\mu} 2^{-\mu}$, where each $a_{\mu}\in\{0, 1\}$.} of $w_k$ and $\theta$, respectively. Now let $v_k$ and $\varphi$ be the rational numbers obtained by truncating the binary expansion of $w_k$, resp. $\theta$, after the $d$-th bit (and keeping the same overall signs). That is, we define $v_k = \mbox{ sign} (w_k)\sum w_{k\mu} 2^{-\mu}$ and $\varphi = \mbox{ sign} (\theta)\sum\theta_{\mu} 2^{-\mu}$ for every $k$, where the sums run from 1 to $d$. Note that \be\label{1-norm_v} \|v\|_1 + |\varphi| \leq \|w\|_1 +|\theta| = 1.\ee It further follows from the definitions of $v$ and $\varphi$ that $|w_k - v_k|\leq 2^{-d}$ and $|\varphi - \theta|\leq 2^{-d}$. This implies that  \be |(v-w)^T \hat x + (\varphi - \theta)|\leq (n+1)2^{-d}.\ee We now choose $d$ to be the smallest positive integer such that $(n+1)2^{-d}$ is strictly smaller than $\epsilon$.
Then, as $|w^T\hat x + \theta|\geq \epsilon$, the quantity
\be\label{eq_margin_weight} w^T\hat x + \theta + \{(v-w)^T \hat x + (\varphi - \theta)\}\ee must have the same sign as $w^T\hat x + \theta$ for every $x$. Note that (\ref{eq_margin_weight}) coincides with $v^T\hat x + \theta$ for every $x$. This shows that $(v, \varphi)$  is also a representation of $f$. But then also multiplying $(v, \varphi)$ with $2^d$ leads to a representation of $f$. The latter representation is integer, and moreover has 1-norm at most $2^d$, since  the 1-norm of $(v, \varphi)$ is at most 1 due to (\ref{1-norm_v}). This shows that the integer weight of $f$ is at most $2^d$. Finally, for our choice of $d$ one has \be\label{equation_d} 2^d \leq 2\cdot \frac{n+1}{\epsilon}.\ee To see this, remark that $d$ is defined to be the smallest integer strictly larger than $\log_2\left[\frac{n+1}{\epsilon}\right]$. But then $d$ must satisfy \be d \leq \log_2\left[\frac{n+1}{\epsilon}\right] + 1,\ee which is equivalent to (\ref{equation_d}). This proves the second inequality in (\ref{margin_int_weight}). \finpr

\section{An equivalent classical computer}\label{sect_class}

Theorem \ref{thm_gaussian} connects the optimal success probability $p$ of computing a linear threshold function $f$ with any matchgate circuit with the margin $\epsilon$ of this LTG. This is a somewhat peculiar connection, e.g. since the definition of the margin of an LTG a priori does not seem to have anything to do with probabilities. In order to understand this relation better, in this section we construct a very simple class of \emph{classical} computers that are capable of (probabilistically) computing LTGs with precisely the same relation between $p$ and $\epsilon$.

To motivate this class, we re-iterate the following simple example which was discussed in section \ref{sect_statement}.  Let $f_{\mbox{\scriptsize{maj}}}$ denote the $n$-bit majority function as before, and let $\epsilon:= n^{-1}$ be its margin. Due to theorem \ref{thm_gaussian}, there exists a matchgate circuit that computes this function with probability $p_{\mbox{\scriptsize{maj}}}$ defined in (\ref{prob_maj}), and this is the optimal success probability which any matchgate circuit can achieve. Now consider the following elementary classical computation, consisting of the following steps:
\begin{itemize}
\item Choose an $n$-bit input string $x$.
\item Generate a random integer  $k$ between 1 and $n$.
\item Output the bit value $x_k$.
\end{itemize}
We now claim that, for every $x$, the output of this computation is $f_{\mbox{\scriptsize{maj}}}(x)$ with probability at least $p_{\mbox{\scriptsize{maj}}}$. To see this, note that the probability $p_x$ that the above procedure outputs 0 is given by \be p_x = \frac{|\{k\in[n]: x_k = 0\}| }{n},\ee and the probability of outputting 1 is $q_x= 1-p_x$. If $x$ contains more zeros than ones---i.e. if $f_{\mbox{\scriptsize{maj}}}(x)$ is zero---then the probability $p_x$ is at least $p_{\mbox{\scriptsize{maj}}}$, as can easily be verified. Similarly, if $x$ contains more ones than zeros, $q_x$ is at least $p_{\mbox{\scriptsize{maj}}}$ as well. Thus, for any $x$ the output of the computation is $f_{\mbox{\scriptsize{maj}}}(x)$ with probability at least $p_{\mbox{\scriptsize{maj}}}$.
In other words, the above classical computation computes the majority function with the same probability of success which can be achieved by the optimal matchgate circuit. Next we will show that every LTG $f$ with margin $\epsilon$ can be associated with a classical computation of the above nature.

The relevant class of classical computations is defined as follows. Fix a probability distribution ${\cal P}:=\{\pi_k\}$ on the set of integers from 1 to $n+1$,  together with a string $c$ of $n+1$ bits. We define the \emph{weighted majority sampling} (WMS) computation  associated with ${\cal P}$ and $c$ to consist of the following steps:
\begin{itemize}
\item Choose an $n$-bit input string $x$.
\item Sample from the distribution ${\cal P}$, yielding   $k\in[n+1]$ with probability $\pi_k$.
\item If $k\leq n$ then output the bit $z_{\mbox{\scriptsize{out}}}:=x_k \oplus c_k$. If $k=n+1$ then output $z_{\mbox{\scriptsize{out}}}:=c_{n+1}$.
\end{itemize}
We say that an $n$-bit boolean function $f$ is WMS computable with probability at least $p$ if there exist ${\cal P}$ and $c$ such that the above three-step procedure yields the output $z_{\mbox{\scriptsize{out}}}=f(x)$ with probability at least $p$ for every $n$-bit input $x$. We prove that the classes of WMS computable and matchgate-computable functions precisely coincide.

\

\noindent {\bf Claim.} {\it Let $p\in(0.5, 1]$. A function is matchgate-computable with probability at least $p$  iff this function is WMS computable with probability at least $p$.}

\

\noindent Consider a WMS computation with associated ${\cal P}$ and $c$. It will be convenient to consider slightly modified computation where the output is $\hat z_{\mbox{\scriptsize{out}}} = (-1)^{z_{\mbox{\scriptsize{out}}}}=\pm 1$ instead of the bit $z_{\mbox{\scriptsize{out}}}$. This will facilitate notation in the proof (but does not play any essential role otherwise). Further, we let $\langle \hat z_{\mbox{\scriptsize{out}}}\rangle_x$ denote the expected value of $\hat z_{\mbox{\scriptsize{out}}}$ given that $x$ is the input of the computation.

We now prove the claim. Let $f$ be an arbitrary $n$-bit boolean function. It follows from section \ref{sect_step1} that $f$ is machgate-computable with probability at least $p$ iff there exists a matchgate circuit $U$ acting on $m\geq n$ qubits such that, for every $x$, one has \begin{itemize}
\item[(a)] $|\langle Z\rangle_x| \geq 2p-1$ and \item[(b)] $\mbox{sign}\langle Z\rangle_x = (-1)^{f(x)}$.\end{itemize}
Furthermore, using an argument analogous to in section \ref{sect_step1}, it can easily be shown that  $f$ is WMS computable with probability at least $p$ if and only if there exist ${\cal P}$ and $c$ such that, for every $x$:
\begin{itemize}
\item[(a')] $|\langle \hat z_{\mbox{\scriptsize{out}}}\rangle_x| \geq 2p-1$ and \item[(b')] $\mbox{sign}\langle \hat z_{\mbox{\scriptsize{out}}}\rangle_x = (-1)^{f(x)}$,\end{itemize} for every $x$.
We thus have to prove that, for every function, conditions (a)-(b) hold for some matchgate circuit $U$ iff (a')-(b') hold for some ${\cal P}$ and $c$.

Suppose first that (a')-(b') are satisfied for some ${\cal P}$ and $c$. We define $w\in \mathbb{R}^n$ and $\theta\in\mathbb{R}$ by \be\label{pi_and_c} w_k:= (-1)^{c_k}\pi_k\quad\mbox{and}\quad \theta:= (-1)^{c_{n+1}} \pi_{n+1},\ee for every $k\in[n]$. Note that the vector $(w, \theta)$ has unit 1-norm. Now let $x$ be an arbitrary bit string and run the WMS computation as described above. Then the expected value of $\hat z_{\mbox{\scriptsize{out}}}$ is
\be\label{hat_z_out'} \langle \hat z_{\mbox{\scriptsize{out}}}\rangle_x = \left\{\sum_{k=1}^{n} \pi_k (-1)^{x_k+ c_k} \right\}+ \pi_{n+1}(-1)^{c_m}= w^T \hat x + \theta.\ee Due to theorem \ref{thm_diag}, there exists a matchgate circuit $U $ on $n+1$ qubits such that \be \langle Z_1\rangle_x := \langle x, 0|U^{\dagger}Z_1U|x, 0\rangle = w^T \hat x + \theta = \langle \hat z_{\mbox{\scriptsize{out}}}\rangle_x.\ee As $\langle Z_1\rangle_x = \langle \hat z_{\mbox{\scriptsize{out}}}\rangle_x$ for every $x$, it follows that conditions (a)-(b) are satisfied for $U$.

To prove the converse, consider an $m$-qubit matchgate circuit $U$ such that (a)-(b) hold. Due to theorem \ref{thm_diag}, there exists  $(v, \varphi)$ with 1-norm at most 1 such that \be \langle Z_1\rangle_x = \langle x, 0| U^{\dagger}Z_1U|x, 0\rangle = v^T\hat x + \varphi\ee for every $n$-bit string $x$. Normalizing $(v, \varphi)$ w.r.t. the 1-norm yields a normalized representation $w:= \gamma v$ and $\theta := \gamma \varphi$, where $\gamma^{-1}:= \|v\|_1 + |\varphi|.$ Note that $\gamma^{-1}\leq 1$. Now choose ${\cal P}$ and $c$ such that (\ref{pi_and_c}) is satisfied. Using an argument similar to the first part of the proof, the expected value of the associated WMS computation is  $\langle \hat z_{\mbox{\scriptsize{out}}}\rangle_x = w^T \hat x + \theta$. We thus have \be \langle \hat z_{\mbox{\scriptsize{out}}}\rangle_x =\gamma (v^T\hat x + \varphi) =\gamma \langle Z_1\rangle_x\ee for every $x$, where $\gamma\geq 1$. Using the identity $\langle \hat z_{\mbox{\scriptsize{out}}}\rangle_x = \gamma \langle Z_1\rangle_x$ and the fact that (a)-(b) hold, it immediately follows that (a') and (b') are satisfied for $({\cal P}, c)$.

\subsubsection*{Acknowledgements}

The author is grateful to S. Bravyi, I. Cirac, R. Jozsa, C. Kraus and K. Vollbrecht for discussions. Work supported by the excellence cluster MAP.

\end{document}